\documentclass[iop]{emulateapj}
\usepackage{apjfonts,amssymb,amsbsy}
\usepackage{graphicx}
\def\apj{ApJ}
\def\apjs{ApJS}

\def\apjl{ApJL}
\def\araa{ARA\&A}

\def\aap{A\&A}
\def\zap{Zeitschrift f\"ur Astrophysik}
\def\mnras{MNRAS}
\def\prl{Phys. Rev. Lett.}
\def\pre{Phys. Rev. E}

\def\lsim{~\raise0.3ex\hbox{$<$}\kern-0.75em{\lower0.65ex\hbox{$\sim$}}~}
\def\gsim{~\raise0.3ex\hbox{$>$}\kern-0.75em{\lower0.65ex\hbox{$\sim$}}~}
 
\slugcomment{ApJL: Draft \today} 
\shorttitle{MOLECULAR CLOUD CONSPIRACY}
\shortauthors{KRITSUK \& NORMAN}

\begin{document}
\title{What Shapes the Structure of Molecular Clouds: Turbulence or Gravity?}

\author{Alexei G. Kritsuk$^\dagger$
and Michael L. Norman$^{\dagger,\;\ddagger}$}
\affiliation{$^\dagger$Physics Department and CASS,
University of California, San Diego; 9500 Gilman Drive, La Jolla, CA 92093-0424, USA\\
$^\ddagger$San Diego Supercomputer Center,
University of California, San Diego;  10100 Hopkins Drive, La Jolla, CA 92093-0505, USA\\
akritsuk@ucsd.edu; mlnorman@sdsc.edu}

\begin{abstract}
We revisit the origin of Larson's scaling relations, which describe the structure and kinematics of molecular
clouds, based on recent observations and large-scale simulations of supersonic turbulence. Using dimensional analysis, 
we first show that both linewidth--size and mass--size correlations observed on scales $0.1-50$~pc can be 
explained by a simple conceptual theory of compressible turbulence without resorting to the often assumed virial 
equilibrium or detailed energy balance condition. The scaling laws can be consistently interpreted as a signature 
of supersonic turbulence with no need to invoke gravity. We then show how self-similarity of structure established 
by the turbulence breaks in star-forming clouds through development of gravitational instabilities in the vicinity 
of the sonic scale, $\ell_{\rm s}\sim0.1$~pc, leading to the formation of prestellar cores. 
\end{abstract}


\section{Introduction}
\citet{larson81} established that for many molecular clouds (MCs) their internal velocity dispersion, $\sigma_{u}$,
is well correlated with the cloud size, $L$, and mass, $m$. Since the power-law form of the correlation,
$\sigma_{u}\propto{}L^{0.38}$, and the power index, $0.38\sim1/3$, were similar to the \citet{kolmogorov41a,kolmogorov41b} 
law of incompressible turbulence (K41), he suggested that the observed nonthermal linewidths may originate from a 
``common hierarchy of interstellar turbulent motions.'' Larson also noticed that the clouds {\em appear} mostly 
gravitationally bound and in approximate virial equilibrium, as there was a close positive correlation between 
the velocity dispersion and mass of the clouds, $\sigma_{u}\propto{}m^{0.20}$, but suggested that these structures 
``cannot have formed by simple gravitational collapse'' and should be at least partly created by 
supersonic turbulence. This seminal paper preconceived many important ideas in the field and
strongly influenced its development for the past 30 years.

\citet{solomon...87} confirmed Larson's study using observations of $^{12}$CO emission with improved 
sensitivity for a large uniform sample of 273 nearby clouds. Their linewidth--size 
relation, $\sigma_{u}=1.0\pm0.1S^{0.5\pm0.05}$~km~s$^{-1}$, however, had a substantially steeper slope
than Larson's, reminiscent of that for clouds in virial equilibrium,\footnote{We deliberately keep the original 
notation used by different authors for the cloud size (e.g., 
the size parameter in parsecs, $S=D\tan(\sqrt{\sigma_l\sigma_b})$; 
the maximum projected linear extent, $L$;
the radius, $R=\sqrt{A/\pi}$, defined for a circle with area, $A$, equivalent to that of cloud) 
to emphasize ambiguity and large systematic errors in the cloud size and mass estimates due 
to possible line-of-sight confusion, ad hoc cloud boundary definitions \citep{heyer...09}, and various X-factors
involved in conversion of a tracer surface brightness into the H$_2$ column density.}   
\begin{equation}
\sigma_{u}=\left(\pi{}G\Sigma\right)^{1/2}R^{1/2},
\label{vir}
\end{equation}
since the \citet{solomon...87} clouds had approximately constant molecular gas surface density, $\Sigma$, 
independent of their radius. The surface density--size 
relation, also known as the third Larson's law, can be derived eliminating $\sigma_u$ from his first two 
relations: $\Sigma\propto{}mL^{-2}\propto{}L^{0.38/0.20-2}\propto{}L^{-0.1}$. Assuming that $\Sigma=const$ 
for all clouds, \citet{solomon...87} evaluated the ``X-factor'' to convert the luminosity in $^{12}$CO~$(1-0)$ 
line to the MC mass. The new power index value $\sim0.5$ ruled out Larson's 
hypothesis that the correlation reflects the Kolmogorov law $\sigma_u\propto{}L^{1/3}$. In the absence 
of  robust predictions for the velocity scaling in supersonic turbulence \citep[cf.][]{passot..88}, 
simple virial equilibrium-based interpretation of linewidth--size relation appealed to many in the 1980s.

Since then views on this subject remain polarized, while it is still uncertain whether gravity 
or turbulence (or magnetic fields) govern the dynamics of MCs. 
For instance, \citet{ballesteros....11,ballesteros.....11} argue that MCs are in a state 
of ``hierarchical and chaotic gravitational collapse,'' while \citet{dobbs..11} believe that
GMCs are ``predominantly gravitationally unbound objects.''

\begin{figure}[t]
\epsscale{1.15}
\plotone{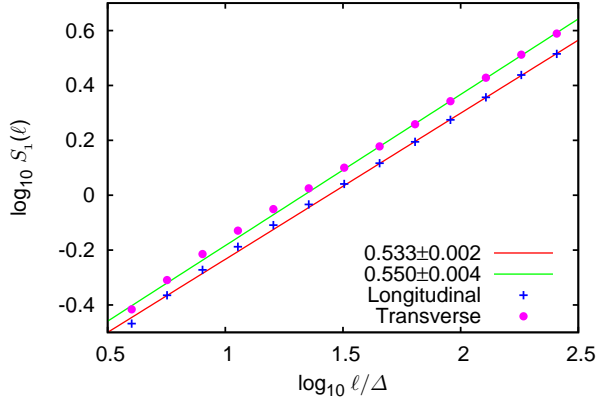}
\caption{Scaling of the first-order transverse (green) and longitudinal (red) velocity SFs 
in a simulation of isothermal supersonic turbulence with $M_{\rm s}=6$ \citep{kritsuk...07a}. 
$\Delta$ is the grid spacing.}
\label{s1}
\end{figure}

\citet{heyer.04} found that the scaling of velocity structure functions (SFs) of 27 GMCs is 
near invariant, 
\begin{equation}
S_1(u, \ell)\equiv\langle\left|\delta{}u_{\ell}\right|\rangle=u_0\ell^{0.56\pm0.02},
\label{hb04}
\end{equation}
for cloudy structures of size $\ell\in[0.03,50]$~pc.\footnote{The lengths entering this relation
are the characteristic scales of the PCA eigenmodes, therefore they may differ from the cloud sizes defined in
other ways \citep{mckee.07}.} In the mean time, numerical simulations of supersonic isothermal turbulence 
returned very similar inertial range scaling exponents for the
first-order velocity SFs ($S_1(u,\ell)\propto\ell^{\zeta_1}$ with $\zeta_1=0.53\pm0.02$ 
and $0.55\pm0.04$ for longitudinal and transverse SFs, respectively, see Fig.~\ref{s1} and 
Kritsuk et al. 2007a). The rms sonic Mach number, $M_{\rm s}=6$, used in these numerical
experiments is characteristic of MC size scale $\ell\approx2$~pc, 
i.e. right in the middle of the observed scaling range. The simulations indicated that the velocity scaling in  
supersonic regimes deviates strongly from Kolmogorov's predictions for fluid turbulence. This result removes one 
of the \citet{solomon...87} arguments against Larson's hypothesis of turbulent origin of the linewidth--size relation. 
Indeed, the scaling exponent of $0.50\pm0.05$ measured by \citet{solomon...87} for whole clouds and a more
recent and precise measurement $0.56\pm0.02$ by \citet{heyer.04} that includes cloud substructure both fall 
right within the range of expected values for supersonic isothermal turbulence at relevant Mach numbers.\footnote{%
Note that the distinction between scaling ``inside clouds'' and ``between clouds'' often used in observational literature
is specious because clouds are not isolated entities on scales of interest \citep[e.g.,][]{henriksen.84}.}

More recently, \citet{heyer...09} used observations of a lower opacity tracer, $^{13}$CO, in a sample of 162 MCs 
with improved angular and spectral resolution to reveal systematic variations of the scaling 
coefficient, $u_0$, in Equation~(\ref{hb04}) with $\ell$ and $\Sigma$. Motivated by the concept
of clouds in self-gravitating equilibrium, they introduced a new scaling coefficient
$u_0^{\prime}\equiv\langle\left|\delta{}u_{\ell}\right|\rangle\ell^{-1/2}\propto\Sigma_{\ell}^{0.5}$.
This correlation would indicate a departure from ``universality'' for the velocity SF
scaling~(\ref{hb04}) and compliance with the virial equilibrium condition~(\ref{vir}). 

\begin{figure}[t]
\epsscale{1.15}
\plotone{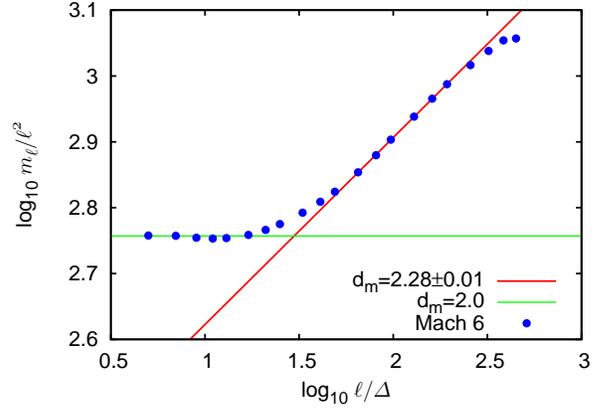}
\caption{Compensated scaling of the mass, $m_{\ell}$, with size $\ell$
in a $2048^3$ simulation of isothermal turbulence with $M_{\rm s}=6$ \citep{kritsuk...09}.
Dissipation-scale structures are shocks with $d_{\rm m}\approx2$. Inertial-range structures have
$d_{\rm m}\approx2.3$  \citep{kritsuk...07a,kritsuk...09}.
}
\label{dm}
\end{figure}

An alternative formulation of the original Larson's third law, $m\propto{}L^{1.9}$, implied a hierarchical
density structure in MCs. Such concept was proposed 60~years ago by von Hoerner to describe a complicated 
statistical mixture of shock waves in highly compressible interstellar turbulence \citep{hoerner51,weizsacker51}. 
Von~Hoerner pictured density fluctuations as a hierarchy of interstellar clouds, analogous to 
eddies in incompressible turbulence. MC observations indeed reveal a pervasive fractal structure 
in the interstellar gas that is interpreted as a signature of turbulence 
\citep{falgarone.91,elmegreen.96,romanduval....10}. The most recent result for a sample of 580 MCs, which 
includes the \citet{solomon...87} clouds, shows a very tight correlation between cloud radii and masses, 
\begin{equation}
m(R)=(228\pm18{\rm M}_{\odot})R^{2.36\pm0.04},\label{rd10}
\end{equation}
for $R\in[0.2,50]$~pc  \citep{romanduval....10}. 
The power-law exponent in this relation is simply the mass dimension of the clouds, 
$d_{\rm m}\approx2.36$, which corresponds to a ``spongy'' medium organized by turbulence into a 
multiscale pattern of clustered corrugated shocks \citep{kritsuk..06}.
Direct measurements of $d_{\rm m}$ in high-resolution three-dimensional simulations of supersonic
turbulence give the inertial sub-range values in excellent agreement with observations: 
$d_{\rm m}=2.39\pm0.01$ \citep[$1024^3$ grid cells,][]{kritsuk...07a} and $2.28\pm0.01$ 
($2048^3$ cells, Fig.~\ref{dm}), depending on details of forcing \citep{kritsuk...10}. 
Note, that $d_{\rm m}\approx2.36$ implies $\Sigma\propto{}mL^{-2}\propto{}L^{0.36}$,
thus the observed mass--size correlation does not support the idea of a universal 
mass surface density of MCs. Meanwhile, positive correlation of $\Sigma$ with $\ell$
removes theoretical objections against the third Larson's law outlined above.

With $d_{\rm m}\approx2.36$, the power-law index in the linewidth--size relation compatible 
with the virial equilibrium condition (\ref{vir}), $\zeta_{1,\rm vir}\equiv(d_{\rm m}-1)/2\approx0.68$, 
is still reasonably close to the scaling exponent $\zeta_1\approx0.56$ in Eq.~(\ref{hb04}), even if one 
assumes $u_0=const$ (see \S3 below). Thus we cannot immediately exclude the possibility of virial equilibrium 
\citep[or kinetic/gravitational energy equipartition, see][]{ballesteros06} across the scale 
range of up to three decades based on the available observations alone. At the same 
time, we have seen that the two classes of observed correlations (e.g., linewidth--size and 
mass--size) are readily reproduced in simulations without self-gravity \citep{kritsuk..11}. 
In this sense, non-gravitating turbulence is self-sufficient at explaining the observations.
Hence, following Occam's razor, it is not necessary to invoke gravity. Meanwhile, in turbulence 
simulations with self-gravity, the velocity power spectra do not show any signature of ongoing
core formation, while the density and column density statistics bear a strong gravitational 
signature on all scales \citep{collins......12}.
What is the nature of this apparent ``conspiracy'' between turbulence and gravity in MCs? 
Why do structures in MCs appear gravitationally bound when they might not be?

In this Letter we use a simple conceptual theory of supersonic isothermal turbulence to show
that scaling exponents in the linewidth--size and mass--size relations are connected, i.e. one
can be derived from the other. In 
\S2 we briefly introduce the concept of compressible cascade and discuss potentially universal 
relations. In \S3 we derive the surface density--size and 
$u_0^{\prime}$--$\Sigma$ relations in several different ways and demonstrate consistency with 
observations and numerical models. \S4 deals with the effects of self-gravity on small scales 
in star-forming clouds, and discusses the origin of the observed mass-size relation for prestellar 
cores. Finally, in \S5 we formulate our conclusions and emphasize the statistical nature of the
observed scaling relations.

\begin{figure}
\epsscale{1.2}
\plotone{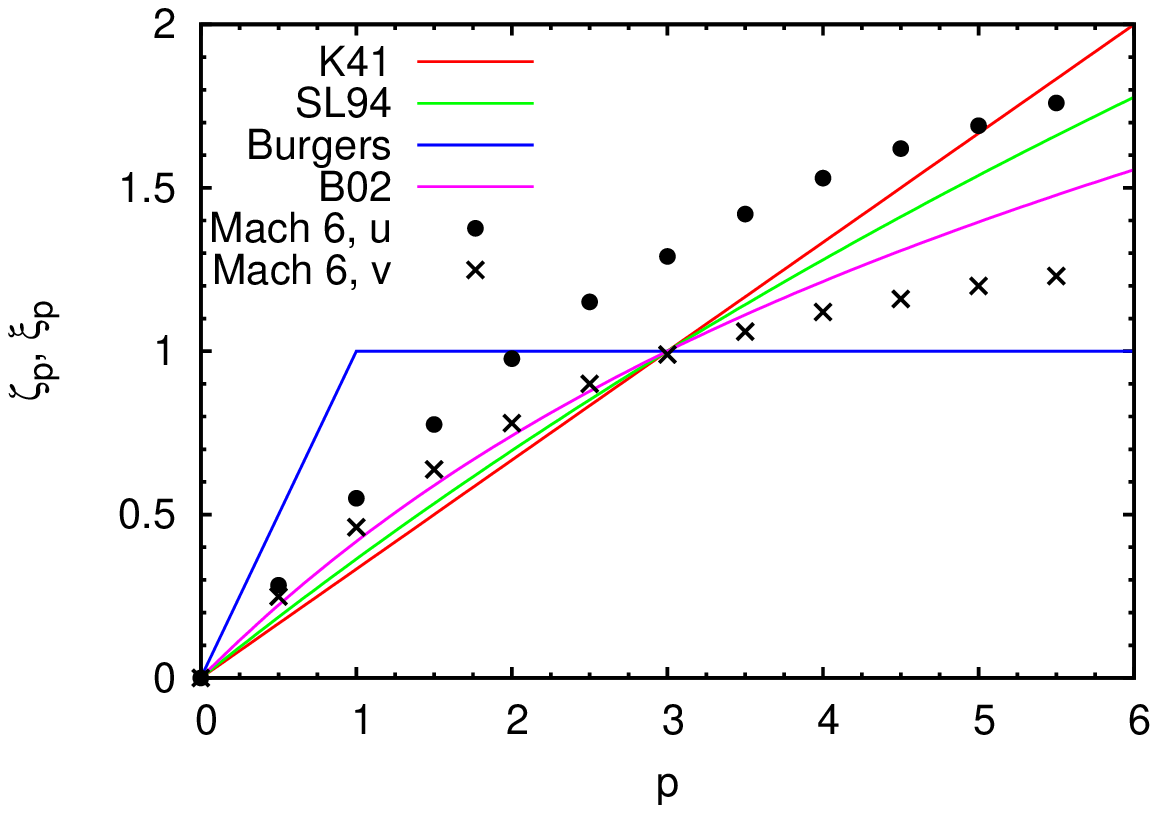}
\caption{Absolute scaling exponents for transverse SFs of velocity ($S_p(u,\ell)\propto\ell^{\zeta_p}$,
circles) and density-weighted velocity $\upsilon=\rho^{1/3}u$ ($S_p(\upsilon,\ell)\propto\ell^{\xi_p}$, crosses) from 
\citet{kritsuk...07a}. Solid lines show $\zeta_p$ predicted by the K41 theory (red), Burgers' model (blue), and 
intermittency models due to \citet[][green]{she.94} and \citet[][magenta]{boldyrev02}. Note that Burgers' model 
predicts $\zeta_1=1$ \citep[cf.][]{mckee.07}.
}
\label{inter}
\end{figure}

\section{What's Universal and What's Not}
In turbulence research, {\em universality} is usually defined as independence on the particular mechanism by
which the turbulence is generated \citep[e.g.,][]{frisch95}. Following this convention, the non-universal nature of 
scaling relation~(\ref{hb04}) can be readily understood. Indeed, any scaling law for compressible turbulence formulated in 
terms of the velocity alone would depend on the Mach number. 
At $M_{\rm s}\lsim1$, density fluctuations are relatively small and turbulence remains very similar to the incompressible
case with $\zeta_1\approx1/3$ \citep{porter..02,benzi.....08}. The scaling steepens at higher Mach
numbers reaching $\zeta_1\approx0.54$ at $M_{\rm s}\approx6$ \citep{kritsuk...07a,kritsuk...07b,pan.11}. 
This transition is also accompanied by a change in the dimensionality of the most singular dissipative structures 
from $1$ (vortex filaments) to $2$ \citep[shocks,][]{boldyrev02,padoan...04}. 

In order to get a universal scaling relation, one would usually have to consider ideal quadratic invariants 
or, more precisely, rugged invariants. In compressible isothermal flows, however, the situation in more complex.
There are two invariants: the mean total energy density $E=\langle\rho{}u^2/2+c_s^2\rho\ln(\rho/\rho_0)\rangle$  
($\rho_0$ is the mean density) and the mean helicity. 
Since $\rho{}u^2/2$ is positive semidefinite, its contribution to $E$ dominates at high Mach numbers, while
the second term is always subdominant. Hence, in the case of MC turbulence, the kinetic energy density is 
probably the best possible choice (if any), as far as universality is concerned. In MCs, on scales above the 
sonic scale, $\ell\gg\ell_{\rm s}\sim0.1$~pc,\footnote{The sonic scale $\ell_{\rm s}$ is defined by the condition 
$\delta{}u_{\ell_{\rm s}}=c_{\rm s}$.} turbulence is highly compressible and any 
quantity with universal scaling must depend on both mass density and velocity.

An illustration is given in Figure~\ref{inter}, where filled circles show the absolute 
scaling exponents, $\zeta_p$, of the velocity SFs of order $p\in[0.5,5.5]$ at $M_{\rm s}=6$. These 
exponents would be close to the K41 prediction, $\zeta_p=p/3$, at $M_{\rm s}\lsim1$, if we assumed isotropy,
homogeneity, and ignored intermittency corrections. At $M_{\rm s}\gsim3$, $\zeta_3$ shows an excess over 
unity, which systematically increases with the Mach number, indicating a non-universal behavior. 
A better candidate for universal scaling is the density-weighted velocity, ${\boldsymbol \upsilon}=\rho^{1/3}{\boldmath u}$ 
since $\xi_3\approx1$ at all Mach numbers \citep{kritsuk...07a,kritsuk...07b}. 
We will use the third-order moment of $\delta\boldsymbol\upsilon$ as its linear 
scaling implies approximately constant kinetic energy flux within the inertial range, 
\begin{equation}
S_3(\upsilon,\ell)\equiv\langle\left|\delta\upsilon_{\ell}\right|^3\rangle=\langle\epsilon\rangle\ell,\label{k07}
\end{equation}
with some minor intermittency correction \citep{kritsuk...07a,aluie11,galtier.11}. Here, $\langle\epsilon\rangle$ 
is the density-weighted energy transfer rate independent of $\ell$. Lower and higher than $p=3$ moments
of $\delta\upsilon$ would most probably require substantial intermittency corrections that are hard to 
predict \citep{landau.44}.
\begin{figure}
\epsscale{1.2}
\plotone{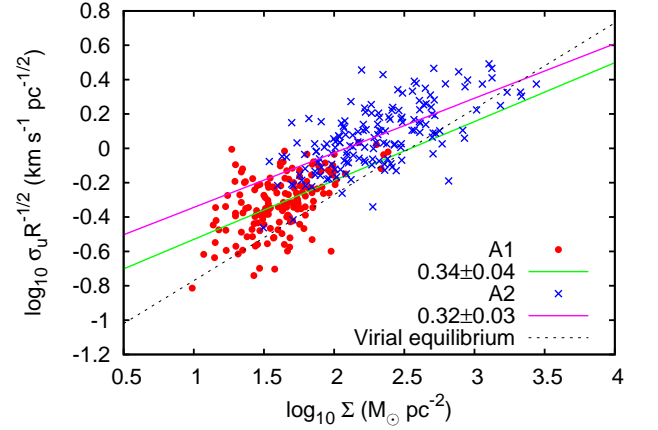}
\caption{Variation of the scaling coefficient $u_0^{\prime}=\sigma_v{}R^{-1/2}$ with mass surface density $\Sigma$
based on data from \citet{heyer...09}. Solid lines with slopes $0.34\pm0.04$ and $0.32\pm0.03$ show least square 
fits to A1 and A2 subsets from \citet{heyer...09}, respectively. Dashed line shows the correlation expected for
clouds in virial equilibrium.
}
\vspace{0.02cm}
\label{sigma}
\end{figure}

\section{Exposing Molecular Cloud Conspiracy}
In the following, we exploit the ``universal'' third-order relation~(\ref{k07}) to derive several secondary 
scaling laws involving $\Sigma_{\ell}$, assuming that the effective kinetic energy flux across the hierarchy 
of scales is approximately constant within the inertial sub-range (and neglecting gravity). 

Dimensionally, the constant spectral energy flux condition,
\begin{equation}
\rho_{\ell}\left(\delta{}u_{\ell}\right)^3\ell^{-1}\propto\Sigma_{\ell}\left(\delta{}u_{\ell}\right)^3\ell^{-2}\propto\Sigma_{\ell}\ell^{3\zeta_1-2}\approx{}const,
\end{equation}
implies
\begin{equation}
\Sigma_{\ell}\propto\ell^{2-3\zeta_1}.
\end{equation}
Substituting $\zeta_1=0.56\pm0.02$, as measured by \citet{heyer.04}, we get $\Sigma_{\ell}\propto\ell^{0.32\pm0.06}$.
We can also rely on the fractal properties of the density distribution to evaluate the scaling of $\Sigma_{\ell}$ with $\ell$: 
$\Sigma_{\ell}\propto\rho_{\ell}\ell\propto{}m_{\ell}/\ell^2\propto\ell^{d_{\rm m}-2}$, which in turn implies 
$\Sigma_{\ell}\propto\ell^{0.36\pm0.04}$ for $d_{\rm m}=2.36\pm0.04$ from \citet{romanduval....10}. Note that both independent estimates for the scaling of $\Sigma_{\ell}$ 
with $\ell$ agree with each other within one sigma. New observational data, thus, indicate that mass surface
density of MCs indeed positively correlates with their size with a scaling exponent $\sim1/3$. This result is
consistent with both observed velocity scaling and observed self-similar structure of the mass distribution in MCs.

Let us now examine data sets A1 and A2 presented in \citet{heyer...09} for a possible correlation of 
$u_0^{\prime}=\sigma_u{}R^{-1/2}$ with $\Sigma=m/\pi{}R^2$. 
Figure~\ref{sigma} shows formal least-square fits for the two data sets with slopes $0.34\pm0.04$ and $0.32\pm0.03$,
respectively. Note that both correlations are not as steep as the virial equilibrium condition~(\ref{vir}) would
imply. When the two data sets are plotted together, however, the apparent shift between A1 and A2 points 
(caused by different cloud boundary definitions in A1 and A2) creates an impression of virial equilibrium condition 
(dashed line in Fig.~\ref{sigma}) being satisfied, although with an offset that \citet{heyer...09} interpret 
as a consequence of LTE-based cloud mass underestimating real masses of the sampled clouds. Each 
of the two data sets, however, suggest scaling with a slope around $1/3$ with larger clouds of higher mass 
surface density being closer to virial equilibrium than smaller structures. 
The same tendency can be traced in the \citet{bolatto....08} sample of extragalactic GMCs \citep[][Fig.~8]{heyer...09}.
A similar trend is recovered by \citet{goodman......09} in the L1448 cloud, where a fraction of self-gravitating material
obtained from dendrogram analysis shows a clear dependence on scale. While most of the emission from the L1448 region
is contained in large-scale bound structures, only a low fraction of small-scale objects appear self-gravitating.

Let us check a different hypothesis, namely whether the observed scaling $\sigma_u{}R^{-1/2}\propto\Sigma^{1/3}$ is
compatible with the turbulent cascade phenomenology and with the observed fractal structure of MCs. The constant
spectral energy flux condition, $\rho_{\ell}\left(\delta{}u_{\ell}\right)^3\ell^{-1}\approx{}const$, can be recast in terms
of $\Sigma_{\ell}\propto\rho_{\ell}\ell$ assuming $\delta{}u_{\ell}\ell^{-1/2}\propto\Sigma_{\ell}^{\alpha}$ with 
$\alpha\approx1/3$ and $\rho_{\ell}\propto\ell^{d_{\rm m}-3}$,
\begin{equation}
\rho_{\ell}\left(\delta{}u_{\ell}\right)^3\ell^{-1}\propto\rho_{\ell}\Sigma_{\ell}\ell^{3/2}\ell^{-1}\propto\ell^{2(d_{\rm m}-3)+3/2}\approx{}const.
\end{equation}
This condition then simply reads as $2(d_{\rm m}-3)+3/2\approx0$ or $d_{\rm m}\approx2.25$, which is
close to the observed fractal dimension of MCs.

As we have shown above, the observed correlation of the scaling coefficient $u_0^{\prime}$ with the coarse-grained
mass surface density of MCs is consistent with a purely turbulent nature of their hierarchical structure and
does not require any additional assumptions concerning virial equilibrium. The origin of this
correlation is rooted in highly compressible nature of the turbulence that implies density
dependence of the lhs of equation~(\ref{k07}). Let us rewrite (\ref{k07}) for the first-order SF
of the density-weighted velocity: $\langle\left|\delta\upsilon_{\ell}\right|\rangle\sim\langle\epsilon_{\ell}^{1/3}\rangle\ell^{1/3}$.
Due to intermittency, the mean cubic root of the dissipation rate is weakly scale-dependent, 
$\langle\epsilon_{\ell}^{1/3}\rangle\propto\ell^{\tau_{1/3}}$, and thus
$\langle\left|\delta\upsilon_{\ell}\right|\rangle\propto\ell^{\xi_1}$, where $\xi_1=1/3+\tau_{1/3}$ and $\tau_{1/3}$ 
is the intermittency correction for the dissipation rate. Using dimensional arguments, 
one can express the scaling coefficient in the \citet{heyer...09} relation,
\begin{equation}
\delta{}u_{\ell}\ell^{-1/2}\propto\rho_{\ell}^{-1/3}\ell^{-1/6+\tau_{1/3}}\propto\Sigma_{\ell}^{-1/3}\ell^{1/6+\tau_{1/3}}.
\end{equation}
Since, as we have shown above, $\Sigma_{\ell}\propto\ell^{1/3}$, one gets
\begin{equation}
\delta{}u_{\ell}\ell^{-1/2}\propto\Sigma^{1/6+3\tau_{1/3}}.
\end{equation}
Numerical experiments give $\tau_{1/3}\approx0.055$ for the density-weighted dissipation rate \citep{pan..09}. 
This value implies scaling, $\delta{}u_{\ell}\ell^{-1/2}\propto\Sigma^{0.33}$, consistent with the \citet{heyer...09} 
data.

\section{A Place for Gravity}
So far, we limited the discussion of Larson's linewidth--size and mass--size relations 
to scales above the sonic scale $\ell_{\rm s}\sim0.1$~pc. Theoretically, the linewidth--size 
scaling index is expected to approach $\zeta_1\approx1/3$ 
at $\ell\lsim\ell_{\rm s}$ in MC sub-structures not affected by self-gravity \citep[see \S2 and][]{kritsuk...07a}. 
\citet{falgarone..09} explored the linewidth--size relation using a large sample of $^{12}$CO 
structures with $\ell\in[10^{-3},10^2]$pc. These data approximately follow a power law 
$\delta{}u_{\ell}\propto\ell^{1/2}$ for $\ell\gsim1$pc. Although the scatter substantially increases 
below 1pc, a slope of $1/3$ ``is not inconsistent with the data.'' $^{12}$CO and $^{13}$CO
observations of translucent clouds indicate that small-scale structures down to a few
hundred AU are possibly intrinsically linked to the formation process of MCs 
\citep{falgarone..98,heithausen04}.

The observed mass--size scaling index, $d_{\rm m}\approx2.36$, is expected to remain constant 
for non-self-gravitating structures down to $\ell_{\eta}\sim30\eta$, which is about a 
few hundred AU, assuming the Kolmogorov scale $\eta\sim10^{14}$~cm \citep{kritsuk+15.11}. 
This trend is traced down to $\sim0.01$pc with recent {\em Herschel} detection of $\sim300$ unbound
starless cores in the Polaris Flare region \citep{andre+10}. For scales below $\ell_{\eta}\sim200$~AU, 
in the turbulence dissipation range, numerical experiments predict convergence to $d_m\approx2$ 
due to shocks, see Fig.~\ref{dm}.

In star-forming clouds, the presence of strongly self-gravitating clumps of high mass surface density
breaks self-similarity imposed by turbulence. One observational signature of gravity 
is the build-up of a high-end power-law tail in the column density PDF associated with filamentary 
structures harboring prestellar cores and YSOs \citep{kainulainen...09,andre...11}.
The power index of the tail, $p=-2/(n-1)$, is determined by the density profile, $\rho\propto{}r^{-n}$, 
of a stable attractive self-similar collapse solution appropriate to the specific conditions 
in the turbulent cloud \citep{kritsuk..11a}. In numerical simulations with self-gravity, we independently 
measured $p\approx2.5$ and $n\approx1.8$ in agreement with the theoretical prediction. This implies 
$d_{\rm m}=3-n\approx1.2$ for the mass--size relation on scales below $\sim0.1$~pc. 
Mapping of the active star-forming Aquila field with {\em Herschel} gave $p=2.7\pm0.1$ and 
$d_{\rm m}=1.13\pm0.07$ for a sample of 541 starless cores with deconvolved FWHM size $\ell\in[0.01,0.1]$~pc 
\citep{konyves+10,andre...11}. Using the above formalism, we get
$d_{\rm m}=3-n=2+2/p\approx1.26$ in reasonable agreement with the direct measurement. Overall, the expected
mass dimension at scales where self-gravity becomes dominant should fall between $d_{\rm m}=1$ 
(Larson-Penston solution, $n=2$) and $d_{\rm m}=9/7\approx1.29$ (pressure-free collapse solution), see 
\citet{kritsuk..11a}.

The characteristic scale where gravity takes control over from turbulence can be predicted using the 
linewidth--size and mass--size relations discussed in previous~\S\S. Indeed, in a turbulent isothermal gas, 
the coarse-grained Jeans mass is a function of scale $\ell$
\begin{displaymath}
m^J_{\ell}\propto\sigma_{\ell}^3\rho_{\ell}^{-1/2}\propto\left\{\begin{array}{ll}\rho_{\ell}^{-1/2}\propto\ell^{(3-d_{\rm m})/2}\propto\ell^{0.32}&\textrm{if $\ell\lsim\ell_{\rm s}$}\\{}\delta\upsilon_{\ell}^3\rho_{\ell}^{-3/2}\propto\ell^{1+3(3-d_{\rm m})/2}\propto\ell^{1.96}&\textrm{if $\ell>\ell_{\rm s}$,}\end{array}\right.
\end{displaymath}
where $\sigma_{\ell}^2\equiv\delta{}u_{\ell}^2+c_{\rm s}^2$ \citep{chandrasekhar51} and we assumed that $\sigma_{\ell}^2\approx{}c_{\rm s}^2$ at $\ell\lsim\ell_{\rm s}$,
while $\sigma_{\ell}^2\approx{}\delta{}u_{\ell}^2$ at $\ell\gsim\ell_{\rm s}$. The dimensionless stability parameter, 
$\mu_{\ell}\equiv{}m_{\ell}/m^J_{\ell}$, shows a strong break in slope at the sonic scale:
\begin{displaymath}\mu_{\ell}\propto\left\{\begin{array}{ll}\ell^{3(d_{\rm m}-1)/2}\propto\ell^{2.04}&\textrm{if $\ell\lsim\ell_{\rm s}$}\\{}\ell^{(5d_{\rm m}-11)/2}\propto\ell^{0.4}&\textrm{if $\ell>\ell_{\rm s}$,}\end{array}\right.
\end{displaymath}
and a rather mild growth above $\ell_{\rm  s}$. Since both $\mu_{\ell}$ and the free-fall time, 
\begin{equation}
t^{\rm ff}_{\ell}\equiv[3\pi/(32G\rho_{\ell})]^{1/2}\propto\ell^{(3-d_{\rm m})/2}\propto\ell^{0.32},
\end{equation}
correlate positively with $\ell$, a {\em bottom-up} nonlinear development of Jeans instability is most likely at $\ell\gsim\ell_{\rm J}$, 
where $\mu_{\ell_{\rm J}}=1$. Note that both $\mu_{\ell}$ and
$t^{\rm ff}_{\ell}$ grow approximately linearly (i.e. relatively weakly) with $\Sigma_{\ell}$ at $\ell>\ell_{\rm s}$, while below the sonic
scale $\mu_{\ell}\propto\Sigma_{\ell}^7$. This means that the instability, if at all present, shuts off rather quickly below 
$\ell_{\rm s}$, i.e. $\ell_{\rm J}\sim\ell_{\rm s}$.
The formation of prestellar cores would be possible only in sufficiently overdense regions on scales around 
$\ell_{\rm s}\sim0.1$~pc. The sonic scale, thus, sets the characteristic mass of the core mass function, $m_{\ell_{\rm s}}$, and the
threshold mass surface density for star formation, $\Sigma_{\ell_{\rm s}}$ \citep[cf.][]{krumholz.05,andre+10}.

\section{Conclusions and Final Remarks}
We have shown that, with current observational data for large samples of Galactic MCs,
a modern version of Larson's relations on scales $0.1-50$~pc can be interpreted as an 
empirical signature of supersonic turbulence fed by the large-scale kinetic energy injection. 
Our interpretation is based on the phenomenology of highly compressible turbulence and 
supported by high-resolution numerical simulations.

Gravity can nevertheless help accumulate the largest molecular structures ($\gsim50$~pc) that appear 
gravitationally bound. 
High-resolution simulations of cloud formation in the general ionized/atomic/molecular turbulent 
ISM context are needed to demonstrate that molecular structures identified as bound in 
position-position-velocity space are indeed genuine three-dimensional objects. On small scales, in 
low-density translucent clouds, 
self-similarity of turbulence can be preserved down to $\sim10^{-3}$~pc, where dissipation starts to become 
important. In contrast, in overdense regions, the formation of prestellar cores breaks the turbulence-induced 
scaling and self-gravity assumes control over the slope of the mass--size relation. We show that the transition 
from turbulence- to gravity-dominated regime in this case occurs close the sonic scale 
$\ell_{\rm s}\sim0.1$~pc, where structures turn gravitationally unstable first, leading to the formation of prestellar 
cores.

Our approach is essentially based on dimensional analysis and the results are valid in a
statistical sense. This means that the scaling relations we discuss hold for averages taken over a large
number of statistically independent realizations of a turbulent flow. Relations obtained for individual MCs and
their internal substructure can show substantial statistical variations around the mean.
The scaling exponents we discuss or derive are usually accurate within $\approx(5-10)$\%, while
scaling coefficients bear substantial systematic errors. Homogeneous multiscale sampling of a large number 
of MCs and their internal structure (including both kinematics and column density mapping) with CCAT, SOFIA
and ALMA will help to detail the emerging picture discussed above.

\acknowledgements
We thank Philippe Andr\'e for providing the scaling exponents based on {\em Herschel} mapping of 
the Aquila field. This research was supported in part by NSF grants AST-0808184, AST-0908740, AST-1109570, 
and by TeraGrid allocation MCA07S014. We utilized computing resources provided by SDSC and NICS.


\begin{thebibliography}{}

\bibitem[Aluie(2011)]{aluie11} Aluie, H.\ 2011, \prl, 106, 174502 

\bibitem[Andr{\'e} et al.(2011)]{andre...11} Andr{\'e}, P., 
Men'shchikov, A., K{\"o}nyves, V., 
\& Arzoumanian, D.\ 2011, Proc. IAU Symp. 270, 255 

\bibitem[Andr{\'e} et 
al.(2010)]{andre+10} Andr{\'e}, P., Men'shchikov, A., Bontemps, S., et al.\ 2010, \aap, 518, L102 

\bibitem[Ballesteros-Paredes et al.(2011b)]{ballesteros.....11}
Ballesteros-Paredes, J., Vazquez-Semadeni, E., Gazol, A., Hartmann, L.~W., 
Heitsch, F., \& Colin, P.\ 2011, \mnras, 416, 1436 

\bibitem[Ballesteros-Paredes et al.(2011a)]{ballesteros....11} 
Ballesteros-Paredes, J., Hartmann, L.~W., V{\'a}zquez-Semadeni, E., 
Heitsch, F., \& Zamora-Avil{\'e}s, M.~A.\ 2011, \mnras, 411, 65 

\bibitem[Ballesteros-Paredes(2006)]{ballesteros06} 
Ballesteros-Paredes, J.\ 2006, \mnras, 372, 443 

\bibitem[Benzi et al.(2008)]{benzi.....08} Benzi, R., Biferale, L., 
Fisher, R.~T., Kadanoff, L.~P., Lamb, D.~Q., 
\& Toschi, F.\ 2008, \prl, 100, 234503 

\bibitem[Bolatto et al.(2008)]{bolatto....08} Bolatto, A.~D., Leroy, 
A.~K., Rosolowsky, E., Walter, F., \& Blitz, L.\ 2008, \apj, 686, 948 

\bibitem[Boldyrev(2002)]{boldyrev02} Boldyrev, S.\ 2002, \apj, 
569, 841 

\bibitem[Chandrasekhar(1951)]{chandrasekhar51} Chandrasekhar, S.\ 1951, 
Roy. Soc. London Proc. Ser. A, 210, 26 

\bibitem[Collins et al.(2012)]{collins......12} Collins, D., Kritsuk, A.~G., Padoan, P.,
Li, H., Xu, H., Ustyugov, S.~D., Norman, M.~L. 2012, ApJ, submitted; arXiv:1202.2594

\bibitem[Dobbs et al.(2011)]{dobbs..11} Dobbs, C.~L., Burkert, 
A., \& Pringle, J.~E.\ 2011, \mnras, 413, 2935 

\bibitem[Elmegreen \& Falgarone(1996)]{elmegreen.96} Elmegreen, 
B.~G., \& Falgarone, E.\ 1996, \apj, 471, 816 

\bibitem[Falgarone et al.(2009)]{falgarone..09} Falgarone, E., Pety, J., \& Hily-Blant, P.\ 2009, \aap, 507, 355 

\bibitem[Falgarone et 
al.(1998)]{falgarone..98} Falgarone, E., Panis, J.-F., Heithausen, A., et al.\ 1998, \aap, 331, 669 

\bibitem[Falgarone 
\& Phillips(1991)]{falgarone.91} Falgarone, E., \& Phillips, T.~G.\ 1991, Fragmentation of Molecular Clouds and Star Formation, 147, 119 

\bibitem[Frisch(1995)]{frisch95} Frisch, U.\ 1995, 
Turbulence.~The legacy of A.~N.~Kolmogorov., Cambridge 
University Press, Cambridge (UK)

\bibitem[Galtier 
\& Banerjee(2011)]{galtier.11} Galtier, S., \& Banerjee, S.\ 2011, \prl, 107, 134501

\bibitem[Goodman et al.(2009)]{goodman......09} Goodman, A.~A., 
Rosolowsky, E.~W., Borkin, M.~A., Foster, J.~B., Halle, M., Kauffmann, J., 
\& Pineda, J.~E.\ 2009, \nat, 457, 63 

\bibitem[Heithausen(2004)]{heithausen04} Heithausen, A.\ 2004, 
\apjl, 606, L13 

\bibitem[Henriksen \& Turner(1984)]{henriksen.84} 
Henriksen, R.~N., \& Turner, B.~E.\ 1984, \apj, 287, 200 

\bibitem[Heyer et al.(2009)]{heyer...09} Heyer, M., Krawczyk, C., 
Duval, J., \& Jackson, J.~M.\ 2009, \apj, 699, 1092 

\bibitem[Heyer 
\& Brunt(2004)]{heyer.04} Heyer, M.~H., \& Brunt, C.~M.\ 2004, \apjl, 615, L45 

\bibitem[Kainulainen et al.(2009)]{kainulainen...09} Kainulainen, J., Beuther, H., Henning, T., \& Plume, R.\ 2009, \aap, 508, L35 

\bibitem[{Kolmogorov(1941a)}]{kolmogorov41a}
Kolmogorov, A.~N. 1941a, Dokl. Akad. Nauk SSSR, 30, 299

\bibitem[{Kolmogorov(1941b)}]{kolmogorov41b}
Kolmogorov, A.~N. 1941b, Dokl. Akad. Nauk SSSR, 32, 19

\bibitem[K{\"o}nyves et 
al.(2010)]{konyves+10} K{\"o}nyves, V., Andr{\'e}, P., Men'shchikov, A., et al.\ 2010, \aap, 518, L106 

\bibitem[Kritsuk et al.(2011c)]{kritsuk+15.11} Kritsuk, A.~G., 
Nordlund, {\AA}., Collins, D., et al.\ 2011c, \apj, 737, 13 

\bibitem[Kritsuk et al.(2011b)]{kritsuk..11} Kritsuk, A.~G., 
Ustyugov, S.~D., \& Norman, M.~L.\ 2011b, Proc. IAU Symp. 270, 179 

\bibitem[Kritsuk et al.(2011a)]{kritsuk..11a} Kritsuk, A.~G., Norman, 
M.~L., \& Wagner, R.\ 2011a, \apjl, 727, L20 

\bibitem[Kritsuk et al.(2010)]{kritsuk...10} Kritsuk, A.~G., 
Ustyugov, S.~D., Norman, M.~L., 
\& Padoan, P.\ 2010, ASP Conf. Ser., 429, 15 

\bibitem[Kritsuk et al.(2009)]{kritsuk...09} Kritsuk, A.~G., 
Ustyugov, S.~D., Norman, M.~L., 
\& Padoan, P.\ 2009, ASP Conf. Ser., 406, 15 

\bibitem[Kritsuk et al.(2007a)]{kritsuk...07a} Kritsuk, A.~G., Norman, 
M.~L., Padoan, P., \& Wagner, R.\ 2007a, \apj, 665, 416 

\bibitem[Kritsuk et al.(2007b)]{kritsuk...07b} Kritsuk, A.~G., Padoan, 
P., Wagner, R., 
\& Norman, M.~L.\ 2007b, AIP Conf. Proc., 932, 393 

\bibitem[Kritsuk et al.(2006)]{kritsuk..06} Kritsuk, A.~G., Norman, 
M.~L., \& Padoan, P.\ 2006, \apjl, 638, L25 

\bibitem[Krumholz 
\& McKee(2005)]{krumholz.05} Krumholz, M.~R., \& McKee, C.~F.\ 2005, \apj, 630, 250 

\bibitem[Landau \& Lifshitz(1944)]{landau.44} Landau, L.~D. \& Lifshitz, E.~M.\ 1944, Mechanics of
Continuous Media, Gostechizdat, Moscow

\bibitem[Larson(1981)]{larson81} Larson, R.~B.\ 1981, \mnras, 
194, 809 

\bibitem[McKee \& Ostriker(2007)]{mckee.07} McKee, C.~F., \& Ostriker, E.~C.\ 2007, \araa, 45, 565 


\bibitem[Padoan et al.(2004)]{padoan...04} Padoan, P., Jimenez, R., 
Nordlund, {\AA}., 
\& Boldyrev, S.\ 2004, \prl, 92, 191102 

\bibitem[Pan 
\& Scannapieco(2011)]{pan.11} Pan, L., \& Scannapieco, E.\ 2011, \pre, 83, 045302 

\bibitem[Pan et al.(2009)]{pan..09} Pan, L., Padoan, P., 
\& Kritsuk, A.~G.\ 2009, \prl, 102, 034501 

\bibitem[Passot et al.(1988)]{passot..88} Passot, T., Pouquet, A., \& Woodward, P.\ 1988, \aap, 197, 228 

\bibitem[Porter et al.(2002)]{porter..02} Porter, D., Pouquet, A., 
\& Woodward, P.\ 2002, \pre, 66, 026301 

\bibitem[Roman-Duval et al.(2010)]{romanduval....10} Roman-Duval, J., 
Jackson, J.~M., Heyer, M., Rathborne, J., 
\& Simon, R.\ 2010, \apj, 723, 492 

\bibitem[She 
\& Leveque(1994)]{she.94} She, Z.-S., \& Leveque, E.\ 1994, \prl, 72, 336 

\bibitem[Solomon et al.(1987)]{solomon...87} Solomon, P.~M., Rivolo, 
A.~R., Barrett, J., \& Yahil, A.\ 1987, \apj, 319, 730 

\bibitem[Vazquez-Semadeni \& Gazol(1995)]{vazquezsemadeni.95} 
Vazquez-Semadeni, E., \& Gazol, A.\ 1995, \aap, 303, 204 

\bibitem[von Hoerner(1951)]{hoerner51} von Hoerner, S.\ 1951, 
\zap, 30, 17 

\bibitem[von Weizs{\"a}cker(1951)]{weizsacker51} von Weizs{\"a}cker, 
C.~F.\ 1951, \apj, 114, 165 

\end{thebibliography}
\end{document}